\documentclass{sig-alternate}

\usepackage{epsfig}
\usepackage{graphicx}
\usepackage{algorithm}
\usepackage{algorithmic}
\usepackage{amssymb}

\usepackage{times,amsmath,epsfig}

\usepackage{textcomp}

\usepackage{subfigure}
\usepackage{algorithmic}
\usepackage{xspace}
\usepackage[normalem]{ulem}
\usepackage{multirow}

\newtheorem{definition}{Definition}
\newtheorem{theorem}{Theorem}

\newtheorem{example}[theorem]{Example}

\begin{document}
%
\title{ELCA Evaluation for Keyword Search on Probabilistic XML Data}

\numberofauthors{2} 

\author{
%
%
%
\alignauthor
Rui Zhou\\
       \affaddr{Faculty of Information and Communication Technologies}\\
       \affaddr{Swinburne University of Technology}\\
       \affaddr{Melbourne, VIC 3122, Australia}\\
       \email{rzhou@swin.edu.au}
\alignauthor 
Chengfei Liu\\
       \affaddr{Faculty of Information and Communication Technologies}\\
       \affaddr{Swinburne University of Technology}\\
       \affaddr{Melbourne, VIC 3122, Australia}\\
       \email{cliu@swin.edu.au}
\and 
\alignauthor
Jianxin Li\\
       \affaddr{Faculty of Information and Communication Technologies}\\
       \affaddr{Swinburne University of Technology}\\
       \affaddr{Melbourne, VIC 3122, Australia}\\
       \email{jianxinli@swin.edu.au}
\alignauthor
Jeffrey Xu Yu\\
       \affaddr{Department of Systems Engineering \& Engineering Management}\\
       \affaddr{The Chinese University of Hong Kong}\\
       \affaddr{Hong Kong, China}\\
       \email{yu@se.cuhk.edu.hk} 
}

\maketitle

\begin{abstract}



As probabilistic data management is becoming one of the main research focuses and keyword search is turning into a more popular query means, it is natural to think how to support keyword queries on probabilistic XML data. With regards to keyword query on deterministic XML documents, ELCA (Exclusive Lowest Common Ancestor) semantics allows more relevant fragments rooted at the ELCAs to appear as results and is more popular compared with other keyword query result semantics (such as SLCAs).

In this paper, we investigate how to evaluate ELCA results for keyword queries on probabilistic XML documents. After defining probabilistic ELCA semantics in terms of possible world semantics, we propose an approach to compute ELCA probabilities without generating possible worlds. Then we develop an efficient stack-based algorithm that can find all probabilistic ELCA results and their ELCA probabilities for a given keyword query on a probabilistic XML document. Finally, we experimentally evaluate the proposed ELCA algorithm and compare it with its SLCA counterpart in aspects of result effectiveness, time and space efficiency, and scalability.

\end{abstract}


\section{Introduction}


Uncertain data management is currently one of the main research focuses in database community. Uncertain data may be generated by different reasons, such as limited observation equipment, unsupervised data integration, conflicting feedbacks. Moreover, uncertainty itself is inherent in nature. This drives the technicians to face the reality and develop specific database solutions to embrace the uncertain world. In many web applications, such as information extraction, a lot of uncertain data are automatically generated by crawlers or mining systems, and most of the time they are from tree-like raw data. In consequence, it is natural to organize the extracted information in a semi-structured way with probabilities attached showing the confidence for the collected information. In addition, dependencies between extracted information can be easily captured by parent-child relationship in a tree-like XML document. As a result, research on probabilistic XML data management is extensively under way. 

Many probabilistic models~\cite{DBLP:conf/pods/SenellartA07,DBLP:conf/vldb/NiermanJ02,DBLP:conf/icde/HungGS03,DBLP:journals/tocl/HungGS07,DBLP:conf/icde/KeulenKA05,DBLP:conf/edbt/AbiteboulS06,DBLP:journals/vldb/AbiteboulKSS09} have been proposed to describe probabilistic XML data. The expressiveness between different models is discussed in~\cite{DBLP:journals/vldb/AbiteboulKSS09}. Beyond the above, querying probabilistic XML data to retrieve useful information is of equal importance. Current studies mainly focused on twig queries~\cite{DBLP:journals/vldb/KimelfeldKS09, DBLP:conf/edbt/ChangYQ09, DBLP:conf/dasfaa/NingLYWL10}, with little light~\cite{DBLP:conf/icde/LiLZW11} shed on keyword queries on probabilistic XML data. However, support for keyword search is important and promising, because users will be relieved from learning complex query languages (such as XPath, XQuery) and are not required to know the schema of the probabilistic XML document. A user only needs to submit a few keywords and the system will automatically find some suitable fragments from the probabilistic XML document.

\begin{figure}[t]
    \centering
    \includegraphics[height=45mm, width=80mm]{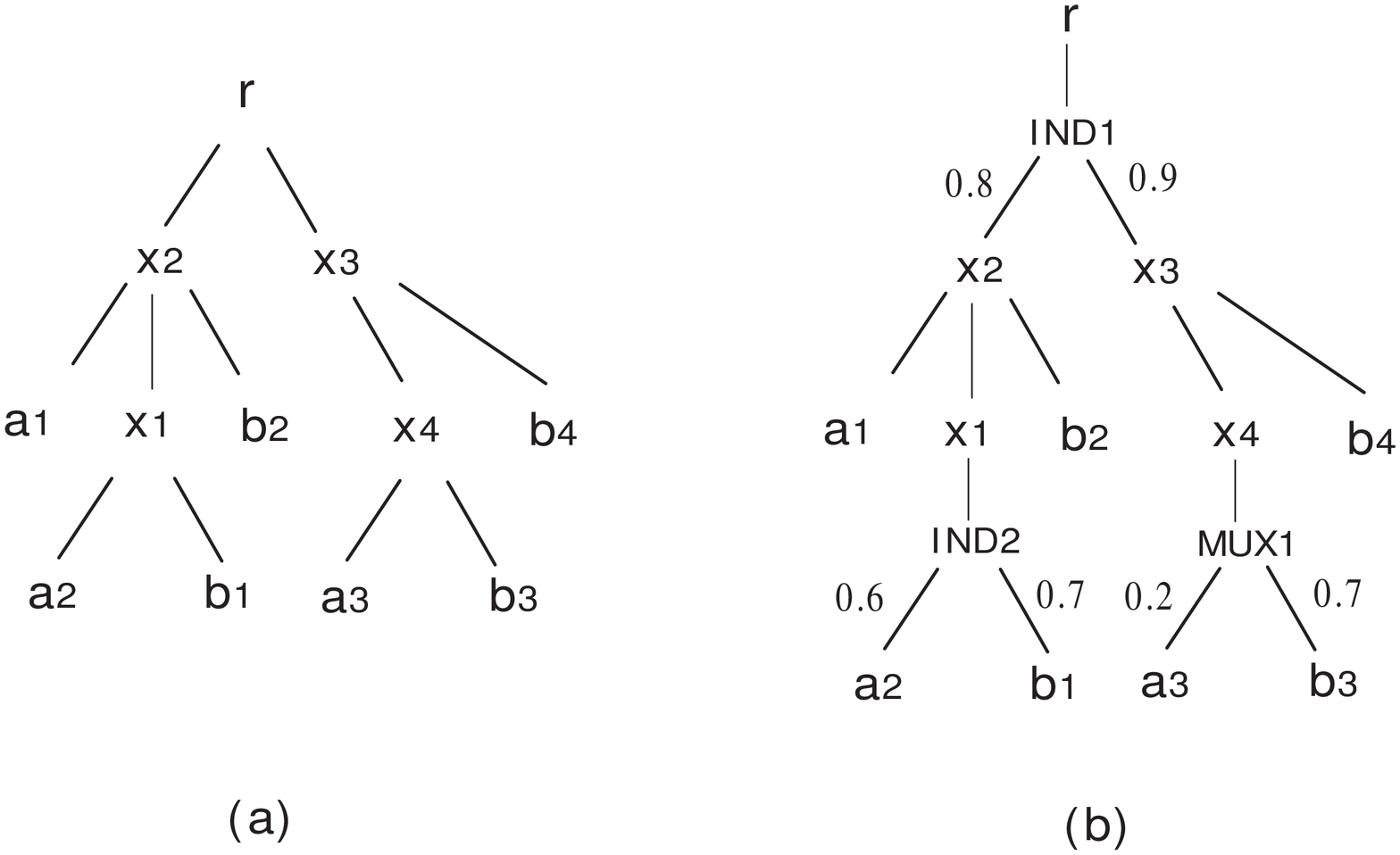}
    \caption{Examples of ELCAs and A Probabilistic XML Tree}
    \label{fig:ELCAexample}
\end{figure}

There has been established works on keyword search over deterministic XML data. One of the most popular semantics to model keyword query results on an deterministic XML document is the ELCA (Exclusive Lowest Common Ancestor) semantics~\cite{DBLP:conf/sigmod/GuoSBS03, DBLP:conf/edbt/XuP08, DBLP:conf/edbt/ZhouLL10}. We introduce the ELCA semantics using an example. Formal definitions will be introduced in Section~\ref{sec:problemdefinition}.  
Fig.~\ref{fig:ELCAexample}(a) shows an ordinary XML tree. Nodes $\{a_1, a_2, a_3\}$ directly contain keyword $a$, and nodes $\{b_1, b_2, b_3, b_4\}$ directly contain keyword $b$. Node $\{x_1, x_2, x_4\}$ are considered as ELCAs of keywords $a$ and $b$. An ELCA is firstly an LCA, and after excluding all its children which contain all keywords, the LCA still contains all the keywords. Node $x_2$ is an ELCA, because after excluding $x_1$ which contains all the keyword, $x_2$ still has its own contributors $a_1$ and $b_2$. Node $x_3$ is not an ELCA, because after excluding $x_4$, $x_3$ only covers keyword $b$. Nodes $x_1$ and $x_4$ are also ELCAs, because they contain both keywords. No children of $x_1$ or $x_4$ contain all the keywords, so no need to exclude any child from $x_1$ or $x_4$. Another popular semantics is SLCA (Smallest LCA) semantics~\cite{DBLP:conf/sigmod/XuP05, DBLP:conf/www/SunCG07}. It asks for the LCAs that are not ancestors of other LCAs. For example, node $x_1$ and $x_4$ are SLCAs on the tree, but $x_2$ is not, because it is an ancestor of $x_1$. It is not difficult to see that the ELCA result is a superset of the SLCA result, so the ELCA semantics can provide more interesting information to users. This motivates us to study the ELCA semantics, and particularly on a new type of data, probabilistic XML data. Note that although SLCA semantics is studied on probabilistic XML data in~\cite{DBLP:conf/icde/LiLZW11}, the solution cannot be used to solve ELCA semantics, as readers may notice that the ELCA semantics is indeed more complex than the SLCA semantics. 

On a probabilistic XML document, nodes may appear or not, accordingly a node is (usually) not certain to be an ELCA. As a result, we want to find not only those possible ELCA nodes, but also their ELCA probabilities. Before we point out the computation challenge, we briefly introduce the probabilistic XML model used throughout this paper. We consider a popular probabilistic XML model,  PrXML$^{\{ind, mux\}}$~\cite{DBLP:conf/vldb/NiermanJ02, DBLP:conf/sigmod/KimelfeldKS08}, where a probabilistic XML document (also called p-document) is regarded as a tree with two types of nodes: \emph{ordinary nodes} and \emph{distributional nodes}. Ordinary nodes store the actual data and distributional nodes define the probability distribution for the child nodes. There are two types of distributional nodes: IND and MUX. IND means the child nodes may appear independently and MUX means the child nodes are mutually-exclusive (i.e. only one child can appear among the defined alternative children). A real number from (0,1] is attached on each edge in the XML tree, indicating the conditional probability that the child node will appear under the parent node given the existence of the parent node. A randomly generated document from a p-document is called a possible world. Apparently, each possible world has a probability. The sum of the probabilities of all possible worlds is 1. A probabilistic XML tree is given in Fig.~\ref{fig:ELCAexample}(b), where unweighted edges have the default probability 1.

Given a keyword query, and a p-document, a node may be an ELCA of the keywords in some possible worlds but not in other possible worlds. We cannot ignore the distributional nodes, because the ELCA results on a probabilistic XML tree may be totally different from those on a deterministic XML tree. For example, in Fig.~\ref{fig:ELCAexample}(b), $x_4$ is no longer an ELCA due to the MUX semantics. $x_3$ may become an ELCA if a possible world contains $a_3$ not $b_3$, but on the deterministic version, $x_3$ is never an ELCA node. Furthermore, $x_1$, a 100\% ELCA node in Fig.~\ref{fig:ELCAexample}(a), becomes a conditional ELCA with probability 0.8*0.6*0.7 in Fig.~\ref{fig:ELCAexample}(b). $x_2$ also becomes an 80\% ELCA node. As a result, deterministic ELCA solutions~\cite{DBLP:conf/sigmod/GuoSBS03, DBLP:conf/edbt/XuP08, DBLP:conf/edbt/ZhouLL10} are not applicable to the new problem. Furthermore, to find out the possible ELCA nodes is not enough. Users may want to know the ELCA probabilities of the possible ELCAs.

To solve the problem, a straightforward and safe method is to generate all possible worlds from the given p-document, evaluate ELCAs using existing ELCA algorithms on deterministic XML for each possible world, and combine the result finally. However, it is obvious that this method is infeasible, because the computation cost is too high, since the number of possible worlds is exponential. The challenge is how to evaluate the ELCA probability of a node using only the p-document without generating possible worlds. 
The idea of our approach is to evaluate the ELCA probabilities in a bottom-up manner.

We summarize the contributions of this paper as follows:
\begin{itemize}

\item To the best of our knowledge, this is the first work that studies ELCA semantics on probabilistic XML data.

\item We have defined probabilistic ELCA semantics for keyword search on probabilistic XML documents. We have proposed an approach on how to evaluate ELCA probabilities without generating possible world and have designed a stack-based algorithm, PrELCA algorithm, to find the probabilistic ELCAs and their probabilities.  

\item We have conducted extensive experiments to test the result effectiveness, time and space efficiency, scalability of the PrELCA algorithm.

\end{itemize}

The rest of this paper is organized as follows. In Section~\ref{sec:problemdefinition}, we introduce ELCA semantics on a deterministic XML document and define probabilistic ELCA semantics on a probabilistic XML document. In Section~\ref{sec:overview}, we propose how to compute ELCA probabilities on a probabilistic XML document without generating possible worlds. An algorithm, PrELCA, is introduced in Section~\ref{sec:algorithms} to explain how to put the conceptual idea in Section~\ref{sec:overview} into procedural computation steps. We report the experiment results in Section~\ref{sec:experimentalresults}. Related works and Conclusion are in Section~\ref{sec:relatedwork} and Section~\ref{sec:conclusions} respectively.

\section{Preliminaries}\label{sec:problemdefinition}

In this section, we first introduce ELCA semantics on a deterministic XML document, and then define probabilistic ELCA semantics on a probabilistic XML document.

\subsection{ELCA Semantics on Deterministic XML}
\label{subsect:ELCA_semantics}

A deterministic XML document is usually modeled as a labeled ordered tree. Each node of the XML tree corresponds to an XML element, an attribute or a text string. The leaf nodes are all text strings. A keyword may appear in element names, attribute names or text strings. If a keyword $k$ appears in the subtree rooted at a node $v$, we say the node $v$ contains keyword $k$. If $k$ appears in the element name or attribute name of $v$, or $k$ appears in the text value of $v$ when $v$ is a text string, we say node $v$ directly contains keyword $k$. A keyword query on a deterministic XML document often asks for an XML node that contains all the keywords, therefore, for large XML documents, indexes are often built to record which nodes directly contain which keywords. For example, for a keyword $k_i$, all nodes directly contain $k_i$ are stored in a list $S_i$ (called inverted list) and can be retrieved altogether at once. 

We adopt the formalized ELCA semantics as the work~\cite{DBLP:conf/edbt/XuP08}. We introduce some notions first. Let $v \prec_a u$ denote $v$ is an ancestor node of $u$, and $v \preceq_a u$ denote $v \prec_a u$ or $v = u$. The function $lca(v_1,\ldots, v_n)$ computes the Lowest Common Ancestor (LCA) of nodes $v_1,\ldots, v_n$. The LCA of sets $S_1,\ldots,S_n$ is the set of LCAs for each combination of nodes in $S_1$ through $S_n$.
\begin{eqnarray*}
\lefteqn{ lca(k_1,\ldots,k_n) = lca(S_1,\ldots,S_n) = } \\
& &  \{ lca(v_1,\ldots,v_n) | v_1 \in S_1,\ldots, v_n \in S_n  \}
\end{eqnarray*}
Given $n$ keywords $\{k_1,\ldots,k_n\}$\, and their corresponding inverted lists $S_1, \ldots, S_n$ of an XML tree $T$, the Exclusive LCA of these keywords on $T$ is defined as:
\begin{eqnarray*}
\lefteqn{ elca(k_1,\ldots,k_n) = elca(S_1,\ldots,S_n) =} \\
& \{ v | \exists v_1 \in S_1, \ldots, v_n \in S_n ( v = lca(v_1,\ldots,v_n) \wedge  \\
& \forall i \in [1,n] \not\exists x ( x \in lca(S_1,\ldots,S_n) \wedge child(v,v_i) \preceq_a x) )  \}
\end{eqnarray*}
where $child(v,v_i)$ denotes the child node of $v$ on the path from $v$ to $v_i$. The meaning of a node $v$ to be an ELCA is: $v$ should contain all the keywords in the subtree rooted at $v$, and after excluding $v$'s children which also contain all the keywords from the subtree, the subtree still contains all the keywords. In other words, for each keyword, node $v$ should have its own keyword contributors. 

\subsection{ELCA Semantics on Probabilistic XML}
\label{subsect:pELCA_semantics}

A probabilistic XML document (p-document) defines a probability distribution over a space of deterministic XML documents. Each deterministic document belonging to this space is called a possible world.
A p-document can be modelled as a labelled tree $T$ with \textit{ordinary} and \textit{distributional} nodes.
Ordinary nodes are regular XML nodes that may appear in deterministic documents, 
while distributional nodes are used for describing a probabilistic process following which possible worlds can be generated. Distributional nodes do not occur in deterministic documents.

We define ELCA semantics on a p-document with the help of possible worlds of the p-document. Given a p-document $T$ and a keyword query $\{k_1, k_2, \ldots, k_n\}$, we define \emph{probabilistic ELCA} of these keywords on $T$ as a set of node and probability pairs $(v,Pr^{G}_{elca}(v))$. Each node $v$ is an ELCA node in at least one possible world generated by $T$, and its probability $Pr^{G}_{elca}(v)$ is the aggregated probability of all possible worlds that have node $v$ as an ELCA. The formal definition of $Pr^{G}_{elca}(v)$ is as follows:
\begin{equation}
Pr^{G}_{elca}(v) = \sum_{i=1}^{m} \{Pr(w_i)|elca(v,w_i)=true\}
\label{eq:defineprob}
\end{equation}   
where $\{ w_1, \ldots, w_m \}$ denotes the set of possible worlds implied by $T$, $elca(v,w_i)=true$ indicates that $v$ is an ELCA in the possible world $w_i$. $Pr(w_i)$ is the existence probability of the possible world $w_i$.

To develop the above discussion, $Pr^{G}_{elca}(v)$ can also be computed with Equation~\ref{eq:globalprob}. Here, $Pr(path_{r \rightarrow v})$ indicates the existence probability of $v$ in the possible worlds. It can be computed by multiplying the conditional probabilities in $T$, along the path from the root $r$ to node $v$. $Pr^{L}_{elca}(v)$ is the local probability for $v$ being an ELCA in $T_{sub}(v)$, where $T_{sub}(v)$ denotes a subtree of $T$ rooted at $v$.
\begin{equation}
Pr^{G}_{elca}(v) = Pr(path_{r \rightarrow v}) \times Pr^{L}_{elca}(v)
\label{eq:globalprob}
\end{equation}
To compute $Pr^{L}_{elca}(v)$, we have the following equation similar to Equation~\ref{eq:defineprob}. 
\begin{equation}
Pr^{L}_{elca}(v) = \sum_{i=1}^{m'} \{Pr(t_i)|elca(v,t_i)=true\}
\label{eq:localprob}
\end{equation}
where deterministic trees $\{t_1, t_2, ..., t_{m'}\}$ are local possible worlds generated from $T_{sub}(v)$, $Pr(t_i)$ is the probability of generating $t_i$ from $T_{sub}(v)$; $elca(v, t_i) = true$ means $v$ is an ELCA node in $t_i$.

In the following sections, we mainly focus on how to compute the local ELCA probability, $Pr^L_{elca}(v)$ for a node $v$. $Pr(path_{r \rightarrow v})$ is easy to obtain if we have index recording the probabilities from the root to node $v$. Then it is not difficult to have the global probability $Pr^G_{elca}(v)$ using Equation~\ref{eq:globalprob}.

\section{ELCA Probability Computation}\label{sec:overview}
In this section, we introduce how to compute ELCA probabilities for nodes on a p-document without generating possible worlds. We start from introducing \emph{keyword distribution probabilities}, and then introduce how to compute the \emph{ELCA probability} for a node $v$ using \emph{keyword distribution probabilities} of $v$'s children.

\subsection{Keyword Distribution Probabilities}\label{subsec:distribution_probability}
Given a keyword query $Q = \{k_1, ... , k_n\}$ with $n$ keywords, for each node $v$ in the p-document $T$, we can assign an array $tab_v$ with size $2^n$ to record the keyword distribution probabilities under $v$. For example, let $\{k_1, k_2\}$ be a keyword query, entry $tab_v[11]$ records the probability when $v$ contains both $k_1$ and $k_2$ in all possible worlds produced by $T_{sub}(v)$; similarly, $tab_v[01]$ stores the probability when $v$ contains only $k_2$; $tab_v[10]$ keeps the probability when $v$ contains only $k_1$; and $tab_v[00]$ records the probability when neither of $k_1$ and $k_2$ appears under $v$. Note that the probabilities stored in $tab_v$ of node $v$ are local probabilities, i.e. these probabilities are based on the condition that node $v$ exists in the possible worlds produced by $T$. To implement $tab_v$, we only need to store non-zero entries of $tab_v$ using a HashMap to save space cost, but, for the clearness of discussion, let us describe $tab_v$ as an array with $2^n$ entries.    

For a leaf node $v$ in $T$, the entries of $tab_v$ are either 1 or 0. Precisely speaking, one entry is ``1'', and all the other entries are ``0''. 
When $v$ is an internal node, let $v$'s children be $\{ c_1, ... , c_m \}$, let $\lambda_i$ be the conditional probability when $c_i$ appears under $v$, then $tab_v$ can be computed using $\{ tab_{c_1}, ... , tab_{c_m} \}$ and $\{ \lambda_1, ... , \lambda_m \}$. We will elaborate the computation for different types of $v$: ordinary nodes, MUX nodes and IND nodes.

\begin{figure*}[t]
    \centering
    \includegraphics[height=110mm, width=170mm]{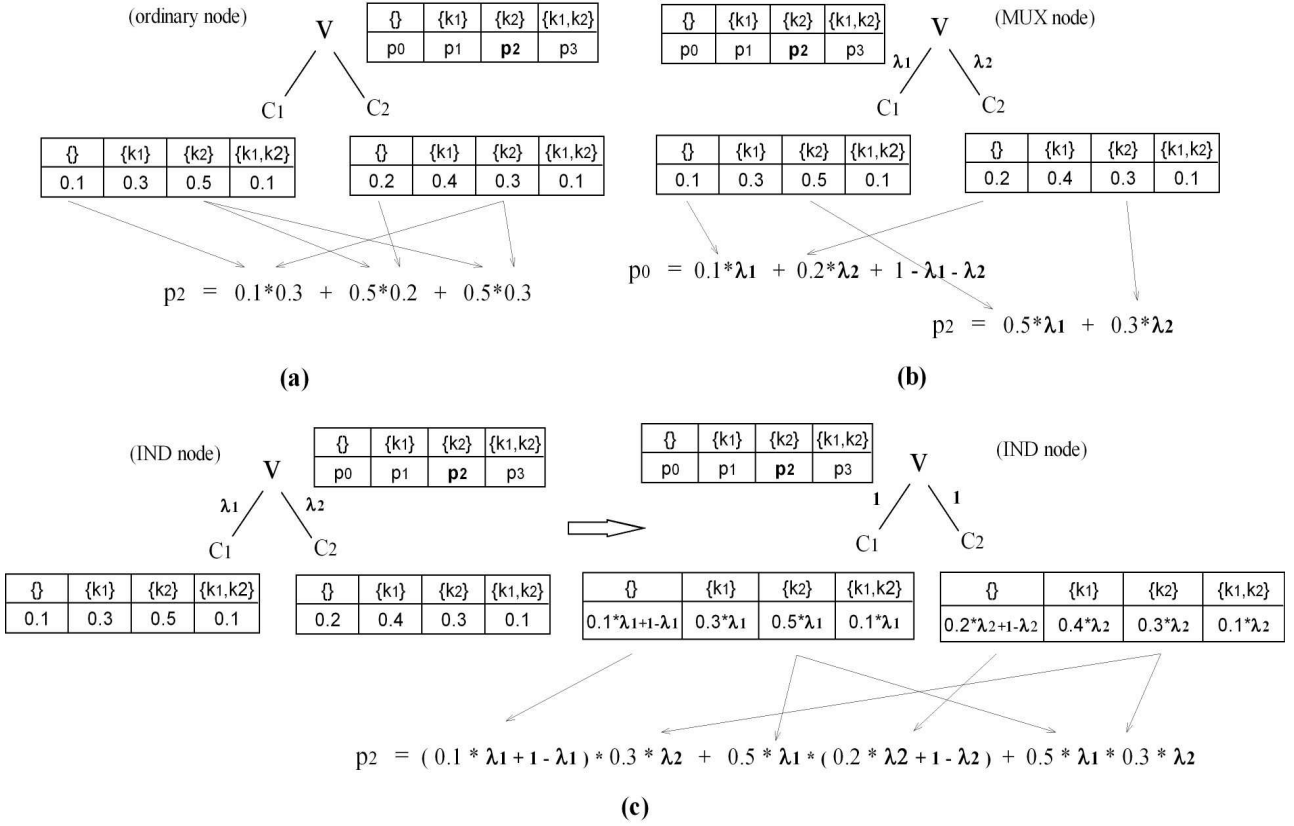}
    \caption{Evaluation of Keyword Distribution Table}
    \label{fig:distributionTable}
\end{figure*}

\begin{figure*}[t]
    \centering
    \includegraphics[height=50mm, width=170mm]{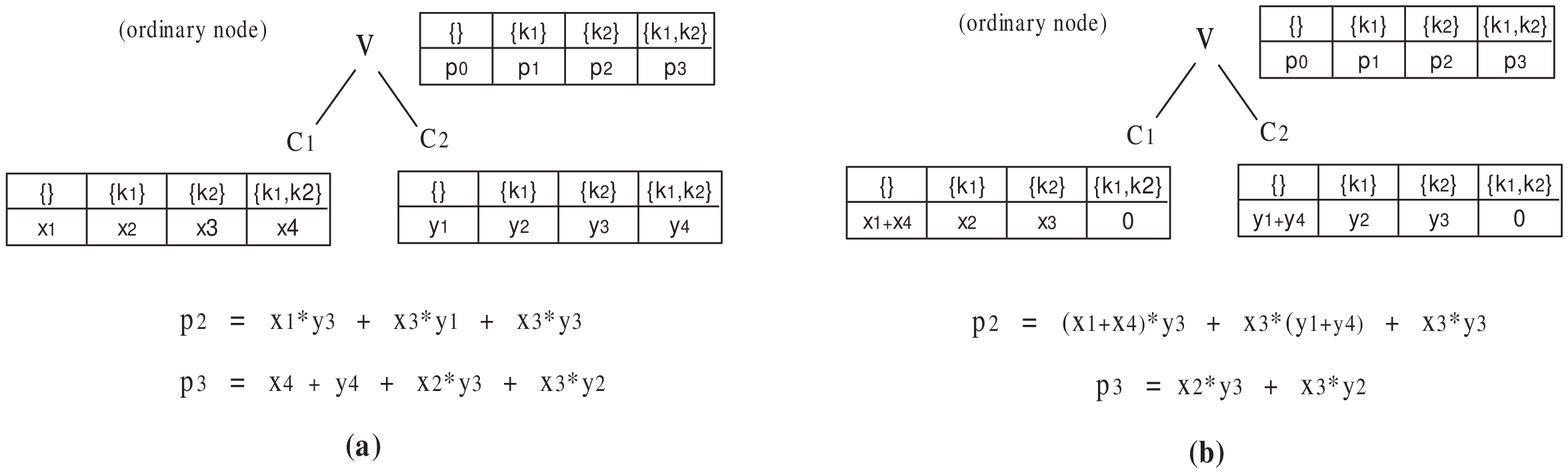}
    \caption{Comparison of keyword distribution probability and ELCA probability}
    \label{fig:distributionTableELCA}
\end{figure*}

\subsubsection{Node $v$ is an Ordinary node} 
\label{subsubsect:ordinary}

When $v$ is an ordinary node, all the children of $v$ will definitely appear under $v$, so we have $ \lambda_1 = ... = \lambda_m = 1$. Let $tab_v[\mu]$ be an entry in $tab_v$, where $\mu$ is a binary expression of the entry index (eg. $tab_v[101]$ refers to $tab_v[5]$, here $\mu$=``101''), then $tab_v[\mu]$ can be computed using the following equation:


\begin{equation}
tab_v [\mu ] \leftarrow \sum\limits_{\mu  = \mu _1  \vee  \ldots  \vee \mu _m } {\prod\limits_{i = 1}^m {tab_{c_i } [\mu _i ]} } 
\label{eq:ordinary_distribution}
\end{equation}
Here, $tab_{c_i}[\mu_i]$ is an entry in $tab_{c_i}$, $\mu_i$ gives the keyword occurrences under $v$'s child $c_i$, and $\mu_1  \vee  \ldots  \vee \mu_m$ gives the keyword occurrences among all $v$'s children. Different $\{ \mu_1, ..., \mu_m \}$ combinations may produce the same $\mu$, so the total probability of these combinations gives $tab_v[\mu]$.

Fig.~\ref{fig:distributionTable} (a) shows an example, where $v$ is an ordinary node. $c_1, c_2$ are $v$'s children. $v$'s keyword distribution table can be computed using $c_1, c_2$'s keyword distribution tables. Take entry $tab_v[01]$, denoted as $p_2$, as an example: $p_2$ stands for the case that $v$ contains keyword $k_2$ but does not contain $k_1$. It correspondingly implies three cases: (1) $c_1$ contains $k_2$ and $c_2$ contains neither $k_1$, $k_2$; (2) $c_2$ contains $k_2$ and $c_1$ contains neither; (3) both $c_1$, $c_2$ only contains keyword $k_2$. The probability sum of the three cases gives the local probability $p_2$.

The naive way to compute $tab_v$ based on Equation~\ref{eq:ordinary_distribution} results in an $O(m2^{nm})$ algorithm, because each $tab_{c_i}$ contains $2^n$ entries, and there are $(2^n)^m = 2^{nm}$ combinations of $\{ \mu_1, ..., \mu_m \}$. For each combination, computing $\prod\nolimits_{i = 1}^m {tab_{c_i } [\mu _i ]}$ takes $O(m)$ time. However, we can compute $tab_v$ progressively in $O(m2^{2n})$ time. The idea is to use an intermediate array $tab'_v$ to record a temporary distribution and then combine the intermediate array $tab'_v$ with each $tab_{c_i}$ \emph{one by one} (\emph{not all together}). We now illustrate the process: at the beginning, $tab'_v$ is initialized using Equation~\ref{eq:ordinary_init}, and then each $tab_{c_i}$ ($i \in [1,m]$) is merged with $tab'_v$ based on Equation~\ref{eq:ordinary_progressive}, in the end, after incorporating all $v$'s children, $tab_v$ is set as $tab'_v$.

\begin{equation}
tab'_v [\mu ] \leftarrow \left\{ {\begin{array}{*{20}c}
   {0 } & {(\mu  \ne 00...0)}  \\
   {1 } & {(\mu  = 00...0) }  \\
\end{array}} \right.
\label{eq:ordinary_init}
\end{equation}

\begin{equation}
tab'_v [\mu ] \leftarrow \sum\limits_{\mu  = \mu ' \vee \mu _i } {tab'_v [\mu '] \cdot tab_{c_i } [\mu _i ]} 
\label{eq:ordinary_progressive}
\end{equation}
 
Generally speaking, we are expecting a keyword query consisting of less than $n \leq 5$ keywords, so $2^{2n}$ is not very large, besides the number of none-zero entries is much smaller than the theoretical bound $2^n$, and therefore we can consider the complexity of computing $tab_v$ as $O(m)$. When we are facing too many keywords, a situation out of the scope of this paper, some preprocessing techniques may be adopted to cut down the number of keywords, such as correlating a few keyword as one phrase. In this paper, we keep our discussion on queries with only a few keywords.

\subsubsection{Node $v$ is an MUX node}

For an MUX node, Equation~\ref{eq:mux_distribution} shows how to compute $tab_v[\mu]$ under mutually-exclusive semantics. A keyword distribution $\mu$ appearing at node $v$ implies that $\mu$ appears at one of $v$'s children, and thus $\sum\nolimits_{i = 1}^m {\lambda _i  \cdot tab_{c_i } [\mu ]}$ gives $tab_v[\mu]$. The case $\mu$=``00...0'' is specially treated. An example is given in Fig.~\ref{fig:distributionTable} (b), consider node $v$ as an MUX node now, then $tab_v[01]$ can be computed as $\lambda_1 \cdot tab_{c_1}[01] + \lambda_2 \cdot tab_{c_2}[01]$. Differently, the entry $tab_v[00]$ includes an extra $(1-\lambda_1-\lambda_2)$ component, because the absence of both $c_1$ and $c_2$ also implies that node $v$ does not contain any keywords.

\begin{equation}
tab_v [\mu ] \leftarrow \left\{ {\begin{array}{*{20}c}
   {\sum\limits_{i = 1}^m {\lambda _i  \cdot tab_{c_i } [\mu ]} } & {(\mu  \ne 00...0)}  \\
   {\sum\limits_{i = 1}^m {\lambda _i  \cdot tab_{c_i } [0]}  + 1 - \sum\limits_{i = 1}^m {\lambda _i } } & {(\mu  = 00...0)}  \\
\end{array}} \right.
\label{eq:mux_distribution}
\end{equation}


Similar to the ordinary case, $tab_v$ can be progressively computed under mutually-exclusive semantics as well. At the beginning, initialize table $tab'_v$ using Equation~\ref{eq:ordinary_init}, the same as the ordinary case; and then $tab'_v$ is increased by merging with $tab_{c_i}$ ($i \in [1,m]$) progressively using Equation~\ref{eq:mux_progressive}. In the end, set $tab_v$ as $tab'_v$. Both the straightforward and the progressive methods take $O(m2^n)$ complexity.

\begin{equation}
tab'_v [\mu ] \leftarrow \left\{ {\begin{array}{*{20}c}
   {tab'_v [\mu ] + \lambda _i  \cdot tab_{c_i } [\mu ] } & {(\mu  \ne 00...0)}  \\
   {tab'_v [0] + \lambda _i  \cdot tab_{c_i} [0] - \lambda_i } & {(\mu  = 00...0)}  \\
\end{array}} \right.
\label{eq:mux_progressive}
\end{equation}

\subsubsection{Node $v$ is an IND node}
When $v$ is an IND node, the computation of $tab_v$ is similar to the ordinary case. Before directly applying Equations~\ref{eq:ordinary_init} and~\ref{eq:ordinary_progressive}, we need to standardize the keyword distribution table. The idea is to transform edge probability $\lambda_i$ into 1, and make corresponding changes to the keyword distribution table with no side-effects. The modification is based on Equation~\ref{eq:ind_transform}. An example is shown in Fig.~\ref{fig:distributionTable} (c), $c_1$ and $c_2$ are two children of IND node $v$ with probabilities $\lambda_1, \lambda_2$, we can equally transform the keyword distribution tables into the right ones and change probabilities on the edges into 1. The $tab_{c_1}[00]$ and $tab_{c_2}[00]$ fields have $(1-\lambda_1)$ and $(1-\lambda_2)$ components, because the absence of a child also implies that no keyword instances could appear under that child. After the transformation, we can compute $v$'s keyword distribution table using the transformed keyword distribution tables of $c_1$ and $c_2$ following the same way as Section~\ref{subsubsect:ordinary}.

\begin{equation}
tab_{c_i} [\mu ] \leftarrow \left\{ {\begin{array}{*{20}c}
   {\lambda _i  \cdot tab_{c_i } [\mu ] } & {(\mu  \ne 00...0)}  \\
   {\lambda _i  \cdot tab_{c_i } [0] + 1 - \lambda _i } & {(\mu  = 00...0)}  \\
\end{array}} \right.\label{eq:ind_transform}
\end{equation}

In summary, we can obtain keyword distribution probabilities for every node in the p-document. The computation can be done in a bottom-up manner progressively. In the next section, we will show how to obtain the ELCA probability of node $v$ using keyword distribution probabilities of $v$'s children.

\subsection{ELCA Probability}
\label{subsect:elca_probability}

We consider ELCA nodes to be ordinary nodes only. We first point out two cases in which we do not need to compute the ELCA probability or we can simply reuse the ELCA probability of a child node, after that we discuss when we need to compute ELCA probabilities and how to do it using keyword distribution table.   

Case 1: $v$ is an ordinary node, and $v$ has a distributional node as a single child. For this case, we do not need to compute ELCA probability for $v$, because the child distributional node will pass its probability upward to $v$. 

Case 2: $v$ is an MUX node and has ELCA probability as 0. According to the MUX semantics, $v$ has a single child. If the child does not contain all the keywords, then $v$ does not contain all the keywords either, on the other hand, if the child contains all the keywords, the child will screen the keywords from contributing upwards. Node $v$ still does not contain its own keyword contributors. In both situations, $v$ is not regarded as an ELCA.

In other cases, including $v$ is an ordinary or IND node and $v$ has a set of ordinary nodes as children, we need to compute ELCA probability for $v$. Note that, when $v$ is an IND node, although $v$ cannot be considered as an ELCA result, we still compute its ELCA probability, because this probability will be passed to $v$'s parent according to Case 1. We discuss the ordinary node case first, IND node is similar. We first define a concept, \emph{contributing distribution}, for the sake of better presenting the idea. 
    
\begin{definition}
Let $\mu$ be a binary expression of an entry index representing a keyword-distribution case, we define $\hat{\mu}$ as the \emph{contributing distribution} of $\mu$ with the value as follows:
\begin{equation}
\hat \mu  \leftarrow \left\{ {\begin{array}{*{20}c}
   \mu  & {(\mu  \ne 11 \ldots 1)}  \\
   {00 \ldots 0} & {(\mu  = 11 \ldots 1)}  \\
\end{array}} \right.
\label{eq:contributing_distribution}
\end{equation}

\label{def:contributing_distribution}
\end{definition}

It means that $\hat{\mu}$ remains the same as $\mu$ in the most cases, except that when $\mu$ is ``11...1'', $\hat{\mu}$ is set to ``00...0''. According to ELCA semantics, if a child $c_i$ of node $v$ has contained all the keywords, $c_i$ will screen the keyword instances from contributing upward to the its parent $v$. This is our motivation to define $\hat{\mu}$. That is to say: when $\mu_i$ is ``11...1'', we regard the contributing distribution $\hat{\mu}_i$ of $\mu_i$ (to parent node $v$) as ``00...0''.

For an ordinary node $v$, let $\{c_1, ..., c_m\}$ be $v$'s children and $\{tab_{c_1}, ..., tab_{c_m} \}$ be the keyword distribution probability arrays of $\{c_1, ..., c_m\}$ respectively, let $\hat{\mu}_i$ be the corresponding contributing distribution of $\mu$, Equation~\ref{eq:ELCA_probability} gives how to compute the local ELCA probability, $Pr_{elca}^L(v)$, for node $v$ using $\{tab_{c_1}, ..., tab_{c_m} \}$. 

\begin{equation}
Pr_{elca}^L (v) \leftarrow \sum\limits_{11...1 = \hat \mu _1  \vee  \ldots  \vee \hat \mu _m } {\prod\limits_{i = 1}^m {tab_{c_i } [\mu _i ]} } 
\label{eq:ELCA_probability}
\end{equation}

To explain Equation~\ref{eq:ELCA_probability}, $v$ is an ELCA when the disjunction of $ \hat{\mu}_1, ..., \hat{\mu}_m $ is ``11...1'', which means after excluding all the children of $v$ containing all the keywords, $v$ still contains all the keywords under other children. All such $\{ \hat{\mu}_1, ..., \hat{\mu}_m \}$ combinations contribute to $Pr_{elca}^L (v)$, and hence the right part of Equation~\ref{eq:ELCA_probability} gives an intuitive way to compute $Pr_{elca}^L (v)$.

Similar to keyword distribution probabilities, we can compute $Pr_{elca}^L (v)$ in a progressive way, reducing the computation complexity from $O(m2^{nm})$ to $O(m2^{2n})$. An intermediate array of size $2^n$ is used, denoted as $tab''_v$. Here, the function of $tab''_v$ is similar to that of $tab'_v$ used in the last section. To be specific, at the beginning, $tab''_v$ is initialized by Equation~\ref{eq:ELCA_progressive_step1}. As the computation goes on, $tab''_v$ is continuously merged with $tab_{c_i}$ ($i \in [1,m]$) using Equation~\ref{eq:ELCA_progressive_step2}. In the end, after merging the intermediate table with all $v$'s children one by one, entry $tab''_v[11...1]$ gives $Pr_{elca}^L (v)$. Note that, although only one entry of $tab''$, $tab''_v[11...1]$, is required as the final result. In the computation, we need to store the whole table $tab''$, because other entries are used to compute the final $tab''_v[11...1]$ entry.

\begin{equation}
tab''_v [\mu ] \leftarrow \left\{ {\begin{array}{*{20}c}
   {0 } & {(\mu  \ne 00...0)}  \\
   {1 } & {(\mu  = 00...0) }  \\
\end{array}} \right.
\label{eq:ELCA_progressive_step1}
\end{equation}

\begin{equation}
tab''_v [\mu ] \leftarrow \sum\limits_{\mu  = \mu ' \vee \hat \mu _i } {tab''_v [\mu '] \cdot tab_{c_i } [\mu _i ]}
\label{eq:ELCA_progressive_step2}
\end{equation}


For each child $c_i$, when we compute $Pr^L_{elca}(v)$, the array entry $tab_{c_i}[11...1]$ acts the same as the entry $tab_{c_i}[00...0]$, because it does not contribute any keyword to its parent. In consequence, we can first modify $tab_{c_i}$ with Equation~\ref{eq:elca_modify}, and reuse Equation~\ref{eq:ordinary_progressive} to compute $Pr^L_{elca}(v)$. 

\begin{equation}
tab_{c_i} [\mu ] \leftarrow \left\{ {\begin{array}{*{20}c}
   {tab_{c_i } [00...0] + tab_{c_i } [11...1] } & {(\mu  = 00...0)}  \\
   {0} & {(\mu  = 11...1)} \\
   {tab_{c_i } [\mu ] } & {otherwise}  \\
\end{array}} \right.\label{eq:elca_modify}
\end{equation}

For an IND node $v$, we can standardize the keyword distribution table using Equation~\ref{eq:ind_transform}. Then, the computation is the same as the ordinary node case.

In Fig.~\ref{fig:distributionTableELCA} (b), we give an example to show how to compute the intermediate table $tab''_v$. An ordinary node $v$ has two children $c_1$, $c_2$. Their keyword distribution tables have been modified according to Equation~\ref{eq:elca_modify}. The probability of $v$ containing both keywords (coming from different children) is given by $p_3 = x_2 \cdot y_3 + x_3 \cdot y_2$, which implies two cases: (1) $c_1$ contains $k_1$ and $c_2$ contains $k_2$; (2) $c_1$ contains $k_2$ and $c_2$ contains $k_1$. Neither $c_1$, $c_2$ are allowed to solely contain both keywords. In ELCA semantics, if a node contains all the keywords, the node will not make contributions to its parent. The probability is smaller than the probability $p_3=x_4 + y_4 + x_2 \cdot y_3 + x_3 \cdot y_2$ (given in Fig.~\ref{fig:distributionTableELCA}(a)), which is the keyword distribution probability when node $v$ contains both keywords, but not required to be from different children. Similarly, the calculation of $tab'_v[01]$ and $tab''_v[01]$ (i.e. $p_2$) are also different. 

\section{Algorithm}\label{sec:algorithms}

In this section, we introduce an algorithm, PrELCA, to put the conceptual idea in the previous section into procedural computation steps. We start with indexing probabilistic XML data, and then introduce PrELCA algorithm, in the end, we discuss why it is reluctant to find effective upper bounds for ELCA probabilities, and it turns out that PrELCA algorithm may be the only acceptable solution.    

\subsection{Indexing Probabilistic XML Data}

\begin{figure}[tbhp]
    \centering
    \includegraphics[height=35mm, width=82mm]{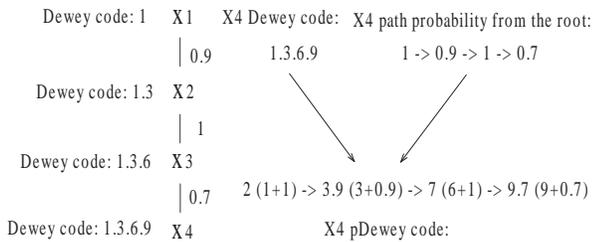}
    \caption{pDewey code}
    \label{fig:deweyIndex}
\end{figure}

We use \emph{Dewey Encoding Scheme}~\cite{DBLP:conf/sigmod/TatarinovVBSSZ02} to encode the probabilistic XML document. By playing a little trick, we can encode edge probability into Dewey code and save some space cost. We illustrate the idea using Fig.~\ref{fig:deweyIndex}. 1.3.6.9 is the Dewey code of the node $x_4$, 0.9->1->0.7 are the probabilities on the path from the root to node $x_4$. To assist with the encoding, we add a dummy probability 1 before 0.9, and get the probability path as 1->0.9->1->0.7. By performing an addition operation, Dewey code and probability can be combined and stored together as 2->3.9->7->9.7. We name the code as pDewey code. For each field $y$ in the combined pDewey code, the corresponding Dewey code can be decoded as $\left\lceil y \right\rceil - 1 $, and the probability can be decoded as $ y + 1 - \left\lceil y \right\rceil$. The correctness can be guaranteed, because edge probabilities always belong to $(0,1]$. Apparently, this encoding trick trades time for space. 

For each keyword, we store a list of nodes that \emph{directly} contain that keyword using B+-tree. The nodes are identified by their pDewey codes. For each node, we also store the node types (ORD, IND, MUX) on the path from the root to the current node. This node type vector helps to perform different types of calculation for different distribution types. For simplicity, we use the traditional Dewey code and omit pDewey code decoding when we introduce the PrELCA algorithm in the next section.

\subsection{PrELCA Algorithm}

According to the probabilistic ELCA semantics (Equation~\ref{eq:defineprob}) defined in Section~\ref{sec:problemdefinition}, a node with non-zero ELCA probability must contain all the keywords in some possible worlds. Therefore, all nodes in the keyword inverted lists and the ancestors of these nodes constitute a candidate ELCA set. The idea of the PrELCA algorithm is to mimic a postorder traversal of the original p-document using only the inverted lists. This can be realized by maintaining a stack. We choose to mimic postorder traversal, because it has the feature that a parent node is always visited after all its children have been visited. This feature exactly fits the idea on how to compute ELCA probability conceptually in Section~\ref{sec:overview}. By scanning all the inverted lists once, PrELCA algorithm can find all nodes with non-zero ELCA probabilities without generating possible worlds. Algorithm~\ref{algo:PrELCA} gives the procedural steps. We first go through the steps, and then give a running example to illustrate the algorithm.

\begin{algorithm}[t]
\caption{PrELCA Algorithm } 
\label{algo:PrELCA}
{  
  \textbf{Input:} inverted lists of all keywords, $S$\\
  \textbf{Output:} a set of $(r[],f)$ pairs $R$, where $r[]$ is a node (represented by its Dewey code), $f$ is the ELCA probability of the node
  \begin{algorithmic}[1]

		\STATE result set $R := \phi$;
		\STATE stack := empty;
    \WHILE{not end of $S$}
    		\STATE Read a new node $v$ from $S$ according to Dewey order, let array $v$[ ] record its Dewey code;

        \STATE $p := lcp(stack, v)$;
        \COMMENT{find the longest common prefix $p$ such that $stack[i].node = v[i]$, $1 \leq i \leq p$}

        \WHILE{$stack.size > p$}
        		\STATE let $r$[] be the Dewey code in the current stack;
        		\STATE let $f = stack.top().elcaTbl[11...1]$;
        		   		
            \STATE add ($r$[], $f$) into the result set R;
            \STATE $popEntry = stack.pop()$;
            \STATE merge $popEntry.disTbl[]$ into $stack.top().disTbl[]$;
            \STATE calculate a new $stack.top().elcaTbl[]$ using the previous $stack.top().elcaTbl[]$ and $popEntry.disTbl[]$;
        \ENDWHILE

        \FOR{$p < j \leq v.length$}
        		\STATE $disTbl[]$ = new disTable();
        		\STATE $elcaTbl[]$ = new elcaTable();
        		\STATE $newEntry = ( node:=v[j]; disTbl[]; elcaTbl[] )$;
            \STATE $stack.push(newEntry)$;

        \ENDFOR
    \ENDWHILE

    \WHILE{$stack$ is not empty}
         \STATE Repeat line 7 to line 13;
    \ENDWHILE
  \end{algorithmic}
}

\end{algorithm}

\begin{figure*}[t]
    \centering
    \includegraphics[height=90mm, width=170mm]{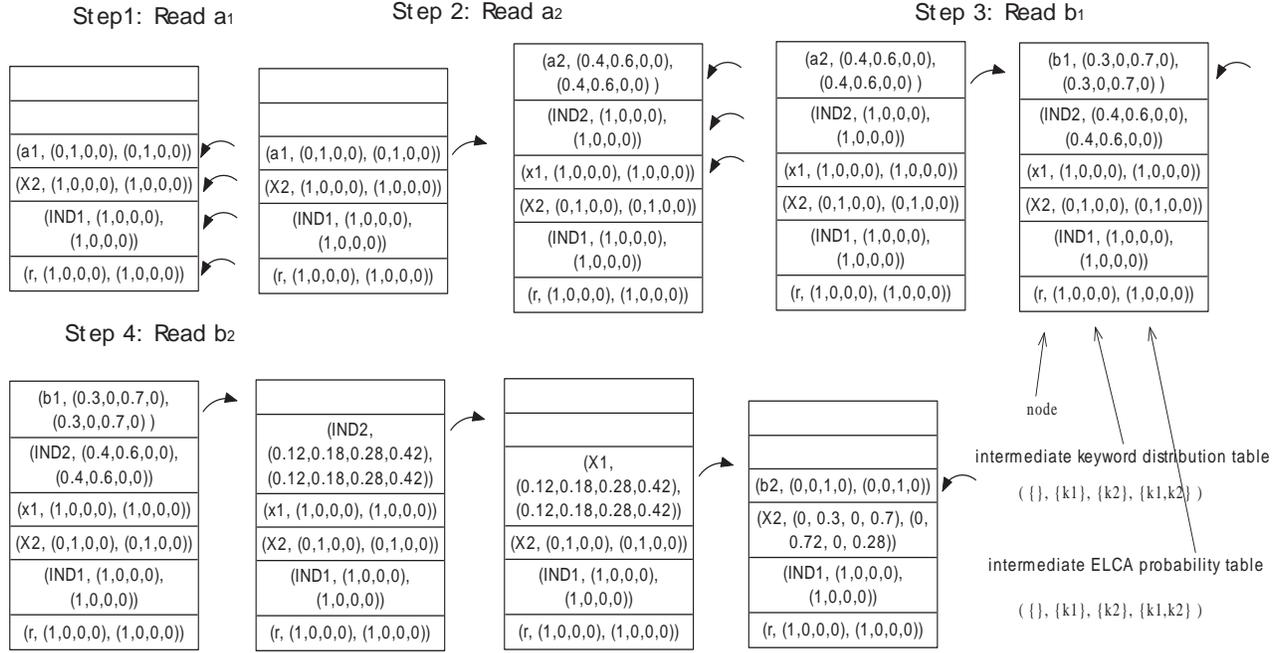}
    \caption{Stack status for some steps of running PrELCA algorithm on the probabilistic XML tree in Fig.2 (b) }
    \label{fig:algorithm}
\end{figure*}

PrELCA algorithm takes keyword inverted lists as input, and outputs all probabilistic ELCA nodes with their ELCA probabilities. The memory cost is a stack. Each entry of the stack contains the following information: (1) a visited node $v$, including the last number of $v$'s Dewey code (eg. 3 is recorded if 1.2.3 is the Dewey code of $v$), the type of the node $v$; (2) an intermediate keyword distribution table of $v$, denoted as $disTbl[]$; (3) an intermediate ELCA probability table of $v$, denoted as $elcaTbl[]$. At the beginning, the result set and the stack are initialized as empty (line 1 and 2). For each new node read from the inverted list (line 3-20), the algorithm will pop up some nodes whose descendant nodes will not be seen in future and output their ELCA probabilities (line 6-13), and push some new nodes into the stack (line 14-19). Line 5 is to calculate how many nodes need to be popped from the stack by finding the longest common prefix between the stack and the Dewey code of the new node. Line 7-9 is to output a result. After that, the top entry will be popped up (line 10), and its keyword distribution table will be merged into the new top entry (which records the parent node of the popped node) based on Equations~\ref{eq:ordinary_progressive} and \ref{eq:mux_progressive} at line 11, and its new top entry's ELCA probability table will also be recalculated based on Equation~\ref{eq:ELCA_progressive_step2} at line 12. For each newly pushed node, its keyword distribution table $disTbl[]$ will be initialized using Equation~\ref{eq:ordinary_init} and Equation~\ref{eq:ELCA_progressive_step1} at line 15 and 16 respectively. Line 17 constructs a stack entry and line 18 pushes the new entry into the stack. After we finish reading the inverted lists, the remaining nodes in the stack are popped and checked finally (line 21-23).

In Fig.~\ref{fig:algorithm}, we show some snapshots for running PrELCA algorithm on the probabilistic XML tree in Fig.~\ref{fig:ELCAexample}(b). At the beginning (step 1), the first keyword instance $a_1$ is read. All the ancestors of $a_1$ are pushed into the stack, with the corresponding $disTbl[]$ and $elcaTbl[]$ fields initialized. 
In step 2, $a_2$ is read according the order of Dewey code. The longest common prefix between the stack and the Dewey code of $a_2$  is $r$.IND1.$x_2$. So $a_1$ is popped up, and $x_2$'s $disTbl[]$ and $elcaTbl[]$ are updated into (0, 1, 0, 0) and (0, 1, 0, 0) by merging with $a_1$'s $disTbl[]$. Node $a_1$ is not a result, because the $a_1$'s $elcaTbl[11]$ is 0. Then, nodes IND2 and $a_2$ are pushed into the stack. In step 3, $b_1$ is read afterwards. Similar to step 2, $a_2$ is popped up with IND2's $disTbl[]$ updated, and then $b_1$ is pushed into the stack. In step 4, we read a new node $b_2$ from the inverted lists. In the stack, node $b_1$ is first popped out of the stack. IND2's disTbl[] is updated into (0.12, 0.18, 0.28, 0.42) by merging $b_1$'s $disTbl[]$ (0.3, 0, 0.7, 0) with IND2's current $disTbl[]$ (0.4, 0.6, 0, 0). Readers may feel free to verify the computation. Similarly, $x_1$'s $disTbl[]$ is updated as (0.12, 0.18, 0.28, 0.42) when IND2 is popped out. $x_1$'s $elcaTbl[]$ is set as IND2's $elcaTbl[]$, because IND2 is a single child distributional node of $x_1$ and thus it does not screen keywords from contributing upwards. (Recall Case 1 in Section~\ref{subsect:elca_probability}). When $x_1$ is popped out, we find $x_1$'s $elcaTbl[11]$ is non-zero. Therefore, $x_1$ has local ELCA probability, $Pr^L_{elca}(x_1)=0.42$. The global ELCA probability for $x_1$ can be obtained by multiplying 0.42 with the edge probabilities along the path from the root $r$ to $x_1$. In this example, the global ELCA probability $Pr^G_{elca}(x_1)= 0.42*0.8$. An interesting scene takes place when $x_1$ is popped out of the stack, $x_2$'s $disTbl[]$ is updated accordingly as (0, 0.3, 0, 0.7) and $x_2$'s $elcaTbl[]$ is updated as (0, 0.72, 0, 0.28). For the first time during the process, $x_2$'s $elcaTbl[]$ is updated into a different value from its $disTbl[]$. The reason is that $x_1$ has screened keyword $a$, $b$ from contributing upwards when $x_1$ itself has already contained both keywords. So the local probability that $x_2$ contains both keywords, represented by $x_2$'s $disTbl[11]$ is 0.7, but the local ELCA probability of $x_2$, represented by $x_2$'s $elcaTbl[11]$ is only 0.28. At last, $b_2$ is pushed into the stack.

\subsection{No Early Stop}

\begin{figure}[tbhp]
    \centering
    \includegraphics[height=25mm, width=50mm]{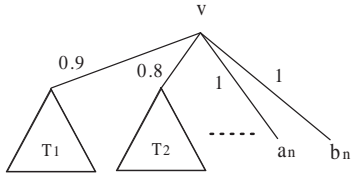}
    \caption{Node $v$ is 100\% an ELCA node, but cannot be discovered until all children have been visited. }
    \label{fig:noearlystop}
\end{figure}

In this subsection, we explain why we need to access all keyword inverted list once, and it is not likely to develop an algorithm that can stop earlier. We use an example to illustrate the idea shown in Fig.~\ref{fig:noearlystop}. Reader can find that node $v$ indeed has the ELCA probability 1, i.e. node $v$ is 100\% an ELCA node, but we are totally unclear about this result when we are examining the previous subtrees $T_1$, $T_2$, etc. One may want to access the nodes in the order of probability values, but it does not change the nature that ELCA probability is always increasing according to Equation~\ref{eq:ELCA_progressive_step2}. Furthermore, that sort of algorithms may need to access the inverted list multiple times, which is not superior compared with the current PrELCA algorithm.

\section{Experiments}\label{sec:experimentalresults}

In this section, we report the performance of the PrELCA algorithm in terms of effectiveness, time and space cost, and scalability. All experiments are done on a laptop with 2.27GHz Intel Pentium 4 CPU and 3GB memory. The operation system is Windows 7, and code is written in Java.

\subsection{Datasets and Queries}

Two real life datasets, DBLP\footnote{http://dblp.uni-trier.de/xml/} and Mondial\footnote{http://www.dbis.informatik.uni-goettingen.de/Mondial/XML}, and one synthetic benchmark dataset, XMark\footnote{http://monetdb.cwi.nl/xml/} have been used. We also generate four test datasets with sizes 10M, 20M, 40M, 80M for XMark data. The three types of datasets are chosen due to the following typical features: DBLP is a large shallow dataset; Modial is a deep, complex, but small dataset; XMark is a balanced dataset, and users can define different depths and sizes to mimic various types of documents. 

For each dataset, we generate a corresponding probabilistic XML tree, using the same method in \cite{DBLP:conf/sigmod/KimelfeldKS08}. To be specific, we traverse the original document in preorder, and for each visited node $v$, we randomly generate some distributional nodes with ``IND'' or ``MUX'' types as children of $v$. Then, for the original children of $v$, we choose some of them to be the children of the new generated distributional nodes and assign random probability distributions to these children with the restriction that the probability sum under a MUX node is no greater than 1. For each dataset, the percentage of the IND and MUX nodes are controlled around 30\% of the total nodes respectively. 
We also randomly select some terms and construct five keyword queries for different datasets, shown in Table~\ref{tab:shortquery}. 

\begin{table}[tbhp] 
  \renewcommand{\arraystretch}{1.3}
  \centering
  \caption{Keyword Queries for Each Dataset}
  \label{tab:shortquery}
    \scalebox{0.8}{
  \begin{tabular}{l|l|l|l}
      \hline\hline
      ID &  Keyword Query & ID &  Keyword Query\\
      \hline
      $X_1$ &  United States, Graduate & $X_2$ &  United States, Credit, Ship \\
      $X_3$ &  Check, Ship  & $X_4$ &  Alexas, Ship  \\ 
      $X_5$ &  Internationally, Ship    &  &    \\
      \hline
      \hline
      $M_1$ &  Muslim, Multiparty   &  $M_2$ &  City, Area\\            
      $M_3$ &  United States, Islands  & $M_4$ &  Government, Area \\
       $M_5$ &  Chinese, Polish     & &\\
      \hline
      \hline
      $D_1$ & Information, Retrieval, Database    & $D_2$ & XML, Keyword, Query  \\             
      $D_3$ & Query, Relational, Database    & $D_4$ & probabilistic, Query       \\        
      $D_5$ & stream, Query      & & \\
      \hline
    \end{tabular}}
\end{table}

In Section~\ref{subsect:effectiveness} and~\ref{subsect:timeandspace}, we will compare PrELCA algorithm with a counterpart algorithm, PrStack~\cite{DBLP:conf/icde/LiLZW11}. We refer PrStack as PrSLCA for the sake of antithesis.
PrStack is an algorithm to find probabilistic SLCA elements from a probabilistic XML document. In Section~\ref{subsect:effectiveness}, we will compare search result confidence (probabilities) under the two semantics. In Section~\ref{subsect:timeandspace}, we will report the run-time performance of both algorithms.

\subsection{Evaluation of Effectiveness}
\label{subsect:effectiveness}

\begin{table}[htbp]
  \renewcommand{\arraystretch}{1.2}
	\small
  \centering
  \caption{Comparison of ELCA and SLCA}
  \label{tab:comparison}
  \begin{tabular}{c|c|c|c|c|c}
    \hline
    \hline
    \multicolumn{2}{c|}{Queries@Mondial} & Max & Min & Avg & Overlap \\[0.5ex]
    \hline
 		\multirow{2}{*}{M1} & ELCA & 0.816  &  0.426  & 0.55  &    \multirow{2}{*}{60\%}     \\ 
    \cline{2-5}
    												& SLCA &   0.703&   0.072 &  0.23 &                               \\
    \hline
    \multirow{2}{*}{M2} & ELCA &  1.000 &  0.980  &  0.99 &    \multirow{2}{*}{100\%}     \\ 
    \cline{2-5}
    												& SLCA &  1.000 &  0.980  &  0.99 &                               \\
    \hline
    \multirow{2}{*}{M3} & ELCA &  0.788 &  0.304  & 0.45  &    \multirow{2}{*}{40\%}     \\ 
    \cline{2-5}
    												& SLCA & 0.582  &  0.073  & 0.13  &                               \\
    \hline
     \multirow{2}{*}{M4} & ELCA &  0.730 &   0.100 &  0.42 &    \multirow{2}{*}{20\%}     \\ 
    \cline{2-5}
    												& SLCA & 0.180  &   0.014 &  0.08 &                               \\
    \hline  
    \multirow{2}{*}{M5} & ELCA & 1.000  & 0.890   & 0.94  &    \multirow{2}{*}{90\%}     \\ 
    \cline{2-5}
    												& SLCA & 1.000  &  0.840  &  0.90 &                               \\
    \hline     
    \hline
    \multicolumn{2}{c|}{Queries@XMark} & Max & Min & Avg & Overlap \\[0.5ex]
    \hline
 		\multirow{2}{*}{X1} & ELCA & 0.560  &  0.165  & 0.27  &    \multirow{2}{*}{20\%}     \\ 
    \cline{2-5}
    												& SLCA &   0.209&   0.054 &  0.15 &                               \\
    \hline
    \multirow{2}{*}{X2} & ELCA &  0.789 &  0.353  &  0.54 &    \multirow{2}{*}{50\%}     \\ 
    \cline{2-5}
    												& SLCA & 0.697 &  0.153  &  0.22 &                               \\
    \hline
    \multirow{2}{*}{X3} & ELCA &  0.970 &  0.553  & 0.62  &    \multirow{2}{*}{30\%}     \\ 
    \cline{2-5}
    												& SLCA & 0.750  &  0.370  & 0.51  &                               \\
    \hline
     \multirow{2}{*}{X4} & ELCA &  0.716 &   0.212 &  0.34 &    \multirow{2}{*}{20\%}     \\ 
    \cline{2-5}
    												& SLCA & 0.236  &   0.014 &  0.13 &                               \\
    \hline  
    \multirow{2}{*}{X5} & ELCA & 0.735  & 0.525   & 0.62  &    \multirow{2}{*}{0\%}     \\ 
    \cline{2-5}
    												& SLCA & 0.163  &  0.044  &  0.08 &                               \\
    \hline
  \end{tabular}
\end{table}

Table~\ref{tab:comparison} shows a comparison of probabilistic ELCA results and probabilistic SLCA results when we run the queries over Mondial dataset and XMark 20MB dataset. For each query and dataset pair, we select top-10 results (with highest probabilities), and record the maximum, the minimum, and average probabilities of the top-10 results. We also count how many results are shared among the results returned by different semantics.  

For some queries, M2 and M5, ELCA results are almost the same as SLCA results (see the Overlap column), but in most cases, ELCA results and SLCA results are different. Query X5 on XMark even returns totally different results for the two semantics. For other queries, at least 20\% results are shared by the two semantics. After examining the returning results, we find that, most of time, PrELCA algorithm will not miss high-ranked results returned by PrSLCA. The reason is that, in an ordinary document, SLCAs are also ELCAs, so probabilistic SLCAs are also probabilistic ELCAs. A node with high SLCA probability is likely to have ELCA probability.

One interesting feature is that, compared with SLCA results, ELCA results always have higher probabilities (except for some queries returning similar results, like M2, M5). For queries M1, M3, M4 on Mondial dataset, the average probability value of ELCA ranges from 0.42 to 0.55, while that of SLCA is about 0.08 - 0.23. On XMark dataset, we have a similar story, with average ELCA probability from 0.27 to 0.62 and average SLCA probability from 0.08 - 0.51. ELCA results also have higher Max and Min values. Since the probability reflects the likelihood that a node exists among all possible worlds as an ELCA or an SLCA, it is desirable that returned results have higher probability (or we say confidence). From this point of view, ELCA results are better that SLCA results, because they have higher existence probabilities. Moreover, the Max probabilities of ELCA results are usually high, above 0.5 in all query cases, but for some queries, such as M4, X5, the Max probabilities of SLCA results are below 0.2. If a user issue a threshold query asking results with probability higher than 0.4, there will be no result using SLCA semantics, but ELCA semantics still gives non-empty results. This could be a reason to use ELCA semantics to return keyword query results.


For the DBLP dataset, we have not listed the results due to paper space limitation, but it is not difficult to understand that probabilistic ELCA results and probabilistic SLCA results are very similar on the DBLP dataset, since it is a flat and shallow dataset.  

\subsection{Evaluation of Time Cost and Space Cost}
\label{subsect:timeandspace}
    
\begin{figure}[t]
  \centering
  \subfigure[Time vs. Query]{\label{fig:querytimexmark}
    \includegraphics[scale=0.62]{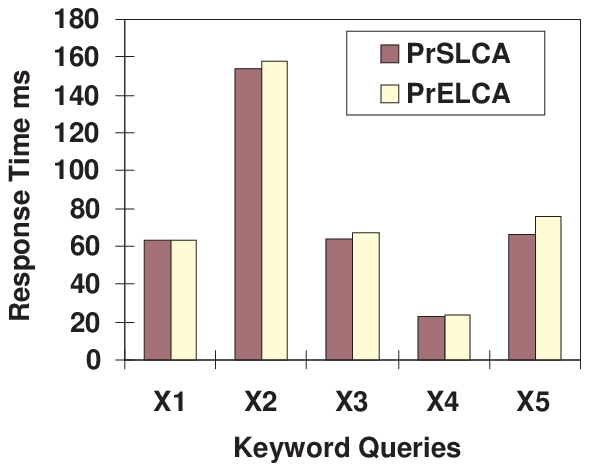}}
  \subfigure[Memory Usage vs. Query]{\label{fig:querymemoryxmark}
    \includegraphics[scale=0.6]{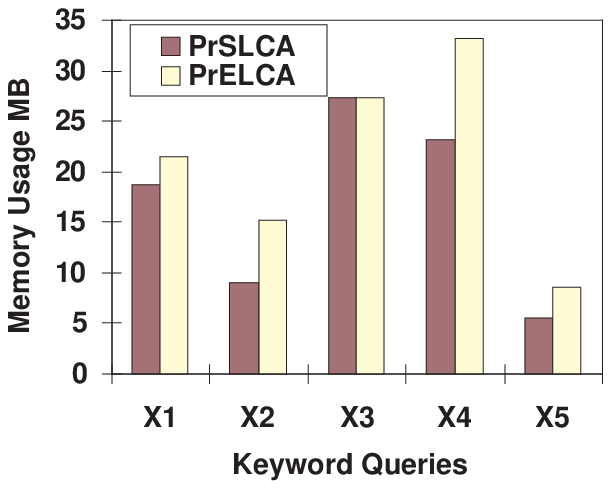}} 
    \\
  \subfigure[Time vs. Query]{\label{fig:querytimemodial}
    \includegraphics[scale=0.6]{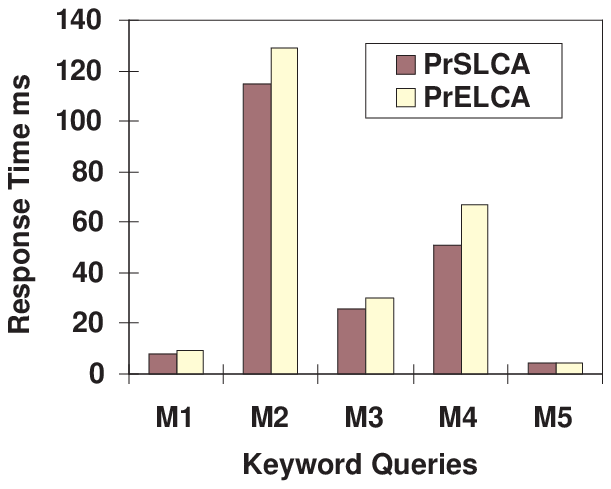}}  
  \subfigure[Memory Usage vs. Query]{\label{fig:querymemorymodial}
    \includegraphics[scale=0.6]{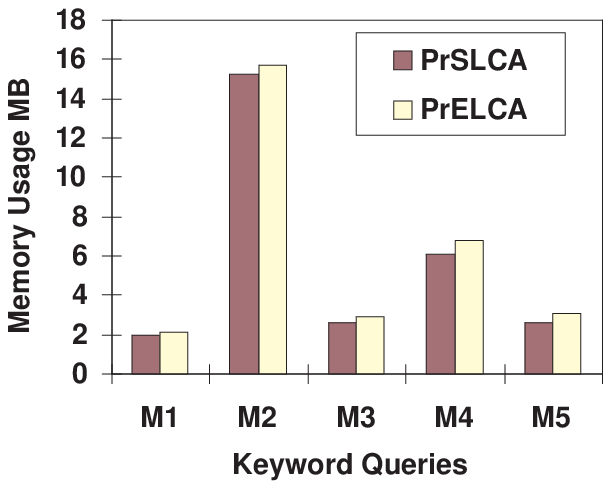}}     
      \\
  \subfigure[Time vs. Query]{\label{fig:querytimedblp}
    \includegraphics[scale=0.6]{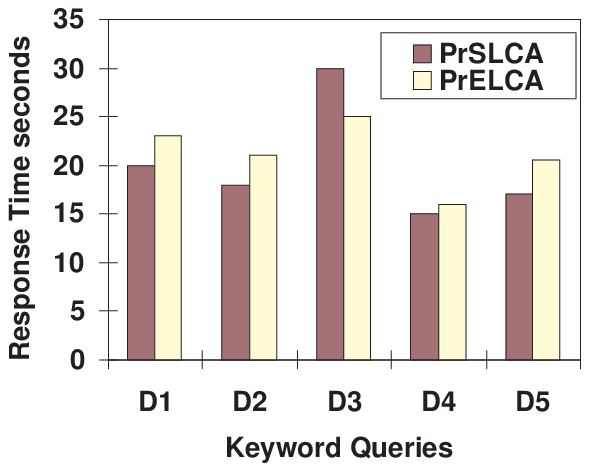}}  
  \subfigure[Memory Usage vs. Query]{\label{fig:querymemorydblp}
    \includegraphics[scale=0.6]{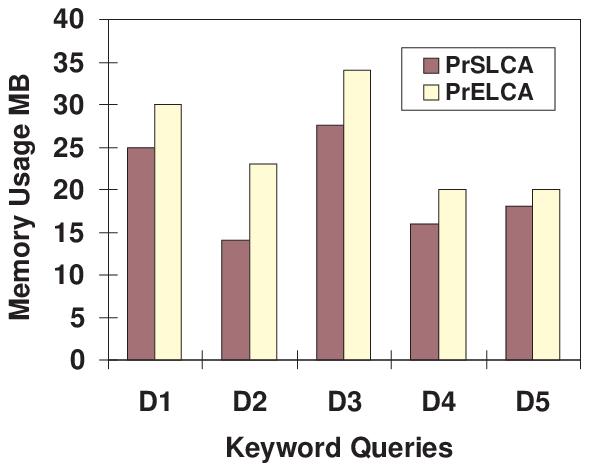}} 
     \caption{Vary Query over Doc2, Doc5, Doc6}
  \label{fig:varyquery} 
\end{figure}

Fig.~\ref{fig:varyquery} shows the time and space cost when we run the queries $X_1$-$X_5$ on Doc2, $M_1$-$M_5$ on Doc5, and $D_1$-$D_5$ on Doc6. From Fig.~\ref{fig:querytimexmark},~\ref{fig:querytimemodial},~\ref{fig:querytimedblp}, we can see that both algorithms PrELCA and PrSLCA are efficient. Although ELCA semantics is more complex than SLCA semantics, PrELCA algorithm has a similar performance as PrSLCA algorithm in terms of time cost. The reason may be that both PrELCA and PrSLCA algorithms are stack-based algorithms and access keyword inverted lists in a similar manner. PrELCA algorithm is slightly slower than PrSLCA in most cases, which is acceptable, because ELCA semantics is more complex and needs more computation. The gap is not large, reflecting that PrELCA algorithm is a competent algorithm if users would like to know probabilistic ELCAs rather than probabilistic SLCAs. From Fig.~\ref{fig:querymemoryxmark}, \ref{fig:querymemorymodial} and~\ref{fig:querymemorydblp}, we can see that PrELCA consumes more memory than PrSLCA. This is because besides the keyword distribution tables which are used in both algorithms, PrELCA has to maintain some other intermediate results to compute the final ELCA probabilities, such as the intermediate table mentioned in Equation~\ref{eq:ELCA_progressive_step1} and~\ref{eq:ELCA_progressive_step2} in Section~\ref{subsect:elca_probability}.
    
\subsection{Evaluation of Scalability}

\begin{figure}[t]
  \centering
   \subfigure[Time vs. Doc. Size]{\label{fig:timedocsizexmark}
    \includegraphics[scale=0.50]{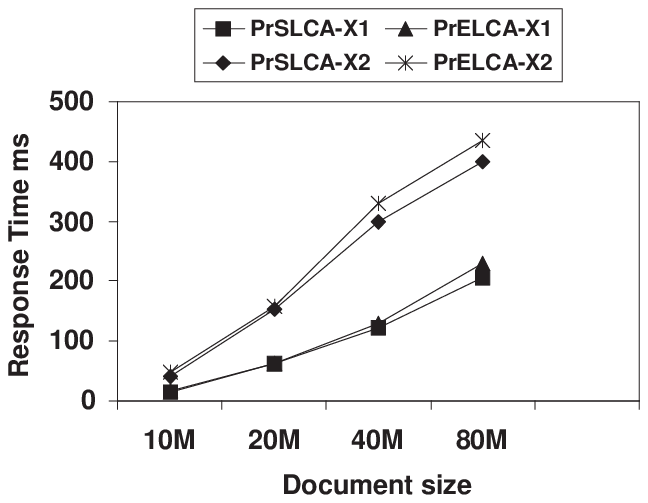}} 
     \subfigure[Memory Usage vs. Doc. Size] {\label{fig:memorydocsizexmark}    
     \includegraphics[scale=0.50]{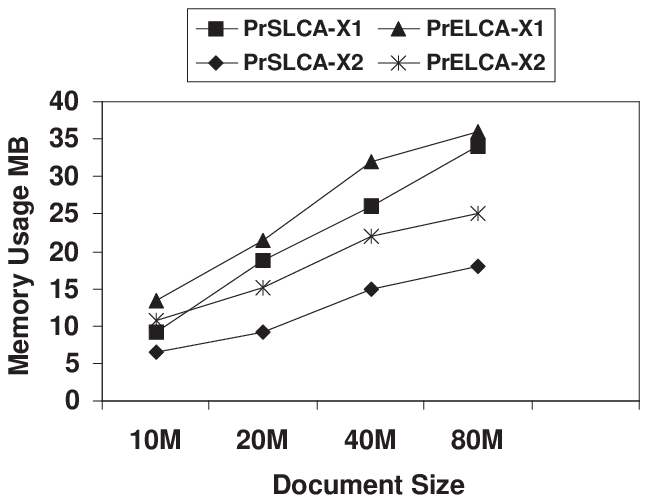}}
  \caption{Vary Document Size}
  \label{fig:documentsize} 
\end{figure}

In this section, we use XMark dataset to test the scalability of the PrELCA algorithm. We test two queries $X_1$, $X_2$ on the XMark dataset ranging from 10M to 80M. Fig.~\ref{fig:timedocsizexmark} shows that the time cost of both queries is going up moderately when the size of the dataset increases. Fig.~\ref{fig:memorydocsizexmark} shows that space cost has a similar trend as the time cost, when document size is increasing. The experiment shows that, for various keyword queries, PrELCA algorithm scales well on different documents, although different queries may consume different memories and run for different time, due to different lengths of the inverted lists.    

\section{Related Work} \label{sec:relatedwork}

There are two streams of works related to our work: probabilistic XML data management and keyword search on ordinary XML documents.

Uncertain data management draws the attention of database research community recently, including both structured and semi-structured data. In the XML context, the first probabilistic XML model is ProTDB~\cite{DBLP:conf/vldb/NiermanJ02}. In ProTDB, two new types of nodes are added into a plain XML document. IND describes independent children and MUX describes mutually-exclusive children. Correspondingly, to answer a twig query on a probabilistic XML document is to find a set of results matching the twig pattern but the results will have existence probabilities. Hung et al.~\cite{DBLP:conf/icde/HungGS03} modeled probabilistic XML documents as directed acyclic graphs, explicitly specifying probability distribution over child nodes. In~\cite{DBLP:journals/tocl/HungGS07}, probabilities are defined as intervals, not points. Keulen et al.~\cite{DBLP:conf/icde/KeulenKA05} introduced how to use probabilistic XML in data integration. Their model is a  simple model, only considering mutually-exclusive sub-elements. Abiteboul and Senellart~\cite{DBLP:conf/edbt/AbiteboulS06} proposed a ``fuzzy trees'' model, where the existence of the nodes in the probabilistic XML document is defined by conjunctive events. They also gave a full complexity analysis of querying and updating on the ``fuzzy trees'' in~\cite{DBLP:conf/pods/SenellartA07}. In~\cite{DBLP:journals/vldb/AbiteboulKSS09}, Abiteboul et al. summarized all the probabilistic XML models
in one framework, and studied the expressiveness and translations between different models. ProTDB is represented as PrXML$^{\{ind, mux\}}$ using their framework. Cohen et al.~\cite{DBLP:journals/tods/CohenKS09} incorporated a set of constraints to express more complex dependencies among the probabilistic data. They also proposed efficient
algorithms to solve the constraint-satisfaction, query evaluation, and sampling problem under a set of constraints. 
On querying probabilistic XML data, twig query evaluation without index (node lists) and with index are considered in~\cite{DBLP:conf/vldb/KimelfeldS07} and~\cite{DBLP:conf/dasfaa/NingLYWL10} respectively. Chang et al.~\cite{DBLP:conf/edbt/ChangYQ09} addressed a more complex situation where result weight is also considered.
The most closest work to ours is~\cite{DBLP:conf/icde/LiLZW11}. Compared to SLCA semantics in~\cite{DBLP:conf/icde/LiLZW11}, we studied a more complex but reasonable semantics, ELCA semantics.

Keyword search on ordinary XML documents has been extensively investigated in the past few years. Keyword search results are usually considered as fragments from the XML document. Most works use LCA (lowest common ancestor) semantics to find a set of fragments. Each fragment contains all the keywords. These semantics include ELCA~\cite{DBLP:conf/sigmod/GuoSBS03, DBLP:conf/edbt/XuP08, DBLP:conf/edbt/ZhouLL10}, SLCA~\cite{DBLP:conf/sigmod/XuP05, DBLP:conf/www/SunCG07}, MLCA~\cite{DBLP:conf/vldb/LiYJ04} and Interconnection Relationship~\cite{DBLP:conf/vldb/CohenMKS03}. Other LCA-based query result semantics rely more or less on SLCA or ELCA by either imposing further conditions on the LCA nodes~\cite{DBLP:conf/cikm/LiFWZ07} or refining the subtrees rooted at the LCA nodes~\cite{DBLP:conf/sigmod/LiuC07, DBLP:journals/pvldb/LiuC08, DBLP:conf/edbt/KongGL09}. 
The works~\cite{DBLP:conf/icde/BaoLCL09} and~\cite{DBLP:conf/edbt/LiLZW10} utilize statistics of the underlying XML data to identify possible query results. All the above works consider deterministic XML trees. Algorithms on deterministic documents cannot be directly used on probabilistic documents, because, on probabilistic XML documents, a node may or may not appear, as a result, a node may be an LCA in one possible world, but not in another. How to compute the LCA probability for a node also comes along as a challenge.

\section{Conclusions}\label{sec:conclusions}

In this paper, we have studied keyword search on probabilistic XML documents. The probabilistic XML data follows a popular probabilistic XML model, PrXML$^{\{ind,mux\}}$. We have defined probabilistic ELCA semantics for a keyword query on a probabilistic XML document in terms of possible world semantics. A stacked-based algorithm, PrELCA, has been proposed to find probabilistic ELCAs and their ELCA probabilities without generating possible worlds. We have conducted extensive experiments to test the performance of the PrELCA algorithm in terms of effectivenss, time and space cost, and scalability. We have compared the results with a previous SLCA based algorithm. The experiments have shown that ELCA semantics gives better keyword queries results with only slight performance sacrifice.

\bibliographystyle{unsrt}
\bibliography{topkslca}  

\end{document}